\documentclass[a4paper,11pt]{article}
\usepackage{pos}
\usepackage{subcaption}
\usepackage{graphicx}

\newcommand{\Frac}{\frac}
\newcommand{\LambQCD}{\Lambda_{QCD}}

\definecolor{db}{rgb}{0.0, 0.0, 0.55}

\title{Three-loop QCD corrections to heavy-to-light \\
	form factors and applications to inclusive $B$ decays}
\ShortTitle{Heavy-to-light form factors to three loops and phenomenological applications}

\author[a,b]{Matteo Fael}
\author[c]{Tobias Huber}
\author[d,e]{Fabian Lange}
\author*[c]{Jakob M\"uller}
\author[f]{Kay Sch\"onwald}
\author[g]{Matthias Steinhauser}

\affiliation[a]{Dipartimento di Fisica e Astronomia “G. Galilei,” Universit\`a di Padova,\\
Via F. Marzolo 8, 35131 Padova, Italy}	
	
\affiliation[b]{Istituto Nazionale di Fisica Nucleare, Sezione di Padova,\\
Via F. Marzolo 8, 35131 Padova, Italy}

\affiliation[c]{Theoretische Physik 1, Center for Particle Physics Siegen (CPPS), Universit\"at Siegen,\\
Walter-Flex-Straße 3, 57068 Siegen, Germany}

\affiliation[d]{Physik-Institut, Universit\"at Z\"urich,\\
Winterthurerstrasse 190, 8057 Z\"urich, Switzerland}

\affiliation[e]{PSI Center for Neutron and Muon Sciences,\\
5232 Villigen PSI, Switzerland}

\affiliation[f]{Theoretical Physics Department, CERN,\\
1211 Geneva, Switzerland}

\affiliation[g]{Institut f\"ur Theoretische Teilchenphysik, Karlsruhe Institute of Technology (KIT),\\
76128 Karlsruhe, Germany}	

\emailAdd{matteo.fael@pd.infn.it}
\emailAdd{fabian.lange@physik.uzh.ch}
\emailAdd{huber@physik.uni-siegen.de}
\emailAdd{jakob.mueller@uni-siegen.de}
\emailAdd{kay.schonwald@cern.ch}
\emailAdd{matthias.steinhauser@kit.edu}

\abstract{We report on the calculation of heavy-to-light form factors at $\mathcal{O}(\alpha_s^3)$ and on selected phenomenological applications in inclusive $B$-decays. After outlining the loop calculation, we extract the hard function in $\bar B \to X_s \gamma$, and discuss our recent progress and preliminary results for the N$^3$LO corrections to partial decay rates in $\bar B \to X_u l \bar \nu_l$, important for the inclusive determination of $|V_{ub}|$. In particular, we establish relations for heavy-quark parameters in the shape-function scheme to four loops and improve a particular model of the $B$ meson shape-function.}

\FullConference{Loops and Legs in Quantum Field Theory (LL2026)\\
12-17, April, 2026\\
Bayreuth, Germany\\}

\begin{document}

\renewcommand{\hookAfterAbstract}{%
	\par\bigskip
	\textsc{SI-HEP-2026-14, P3H-26-051, ZU-TH 23/26, CERN-TH-2026-159, TTP26-026}
}

\maketitle

\section{Introduction}
In many particle physics processes form factors are important ingredients for calculating phy-\\sical observables. A particularly interesting class consists of heavy-to-light form factors, appearing e.g.\ in inclusive $B$-meson decays and top-quark physics.
In particular in $B$-physics, assuming the light fermion to be massless suffices for (most) phenomenological applications. While these form factors have been known for several years at NNLO~\cite{Bonciani:2008wf,Asatrian:2008uk,Beneke:2008ei,Bell:2008ws,Huber:2009se,Bell:2010mg}, only the colour-planar corrections to the (pseudo-)scalar and (axial-)vector form factors were calculated at N$^3$LO~\cite{Chen:2018fwb,Datta:2023otd}. Also the structure functions at NNLO~\cite{Broggio:2026edk} and N$^3$LO~\cite{Chen:2026jwl} have become available earlier this year. In the recent publication~\cite{Fael:2024vko}, we calculate the heavy-to-light form factors for all possible current insertions in full QCD to three loops, still neglecting the fermion mass. \\
We then use our results for the form factors to calculate the finite hard matching coefficients at leading power in the framework of Soft-Collinear Effective Theory (SCET), which allows us to extract the hard function in the factorization theorem for the photon energy spectrum in inclusive $\bar B \to X_s \gamma$ at N$^3$LO. \\
As another application of the form factors we investigate partial decay rates in $\bar B \to X_u l \bar \nu_l$ decays at N$^3$LO within the Bosch-Lange-Neubert-Paz (BLNP)~\cite{Bosch:2004th,Lange:2005yw} framework. The form factors constitute one of the last missing pieces for investigating these inclusive decays at third order in perturbation theory, important for a precise inclusive extraction of the Cabibbo–Kobayashi–Maskawa matrix element $|V_{ub}|$. The current values of the inclusive and exclusive determination are~\cite{Belle-II:2025pye}
\begin{align}\label{eq::vubExclIncl}
	|V_{ub}|^{\rm excl.} = (3.43 \pm 0.12) \times 10^{-3}\,, |V_{ub}|^{\rm incl.} = (4.06 \pm 0.16) \times 10^{-3}\,,
\end{align}
and hence differ by about three standard deviations. We report on our ongoing efforts to examine the effect of the third-order corrections on the determination of $|V_{ub}|$. \\
The remainder of the text is organized as follows: In Section \ref{sec::generalloopcalc}, we review and outline the loop calculation, Section \ref{sec::hardbsg} is devoted to the hard function in $\bar B \to X_s \gamma$. In Section \ref{sec::inclvub}, we show our recent results on the partial rates in $\bar B \to X_u l \bar \nu_l$ decays at N$^3$LO and conclude in Section \ref{sec::conclusion}.
\section{Heavy-to-light form factors to three loops}\label{sec::generalloopcalc}
For the calculation of the heavy-to-light form factors we set up the amplitude in QCD for generic external scalar, pseudoscalar, vector, axialvector and tensor current. We assume one on-shell massless quark with momentum $q_1$ and one on-shell massive quark with momentum $q_2$ and mass $m$. The momentum transfer is $s \equiv q^2 = (q_1-q_2)^2$. \\
The calculation of the heavy-to-light form factors is performed in multiple steps typical for a modern multi-loop calculation. In Ref.~\cite{Fael:2024vko}, we discuss all steps in more detail and highlight here only the main points of the calculation.
The general calculation for $s \neq 0$ is done via the method of projectors in general $R_\xi$-gauge. We employ the tools {\tt qgraf}~\cite{Nogueira:1991ex}, {\tt tapir}~\cite{Gerlach:2022qnc}, {\tt exp}~\cite{Harlander:1998cmq,Seidensticker:1999bb}, the in-house {\tt FORM} code {\tt calc}, {\tt Kira}~\cite{Maierhofer:2017gsa,Klappert:2020nbg,Lange:2025fba} and {\tt FireFly}~\cite{Klappert:2019emp,Klappert:2020aqs}. For the Integration-by-Parts (IBP) reduction using {\tt Kira} we are trying to find a basis of master integrals such that the dependence on $d=4 -2 \epsilon$ and $s/m^2$ is factorized in the denominators of the coefficients in the IBP tables and the number of spurious poles in $\epsilon$ is reduced~\cite{Smirnov:2020quc,Usovitsch:2020jrk}. We first perform reductions for sample integrals in each integral family to find all master integrals. Then, we employ an improved version of {\tt ImproveMasters.m}~\cite{Smirnov:2020quc} to find a more suitable basis. After that, we perform the full reduction using the so-obtained master integrals and do a final run of all families together to detect further symmetries among the master integrals, yielding $429$ master integrals in total. We further carry out an independent calculation of the form factors at $s=0$ for the tensor current important for $\bar B \to X_s \gamma$ using tensor reduction and Feynman gauge. In this case the tool chain is {\tt qgraf}, an in-house {\tt FORM} code, {\tt FeynHelpers} \cite{Shtabovenko:2016whf}, {\tt Feynson}~\cite{Maheria:2022dsq}, {\tt Kira} and {\tt FireFly}. We obtain $246$ master integrals in this instance.
\subsection{Master integrals}
\label{sec::mis}
To evaluate the master integrals, we set up differential equations in $x=s/m^2$ using {\tt LiteRed}~\cite{Lee:2013mka}, and use {\tt Kira} and {\tt FireFly} for the subsequent IBP reductions. We consider all master integrals at one- and two-loop level analytically. At three loops we calculate all master integrals related to colour-planar topologies, topologies depending on the number of light flavours $n_l$$(=4)$ and those with two heavy-fermion loops analytically. We achieve this by using methods of Ref.~\cite{Ablinger:2018zwz}, employing {\tt Sigma}~\cite{sigmaII} and {\tt HarmonicSums}~\cite{Ablinger:2018cja}. More specifically, we do not try to find a canonical basis of master integrals, but we uncouple blocks of the differential equation into higher-order ones and solve these via factorization of the differential operator and variation of constant. This is possible because we can factorize the differential operators to first order and thus can express the results as iterated integrals over the algebraic letters $1/x,1/(1\pm x),1/(2-x)$. In the full amplitude elliptic sectors emerge, hence we cannot use the discussed method for all master integrals. We fix the boundary conditions either by direct integrations, Mellin-Barnes techniques, regularity conditions or {\tt PSLQ} algorithm on numerical results obtained from {\tt AMFlow}~\cite{Liu:2022chg}. \\
We compute the remaining master integrals semi-analytically via the method "Expand-and-\\
Match"~\cite{Fael:2021kyg,Fael:2022rgm,Fael:2022miw,Fael:2023zqr}. In essence, we do series expansions about regular and singular points of the differential equations and numerically match neighbouring expansions where both converge. As expansion points we choose $x \!= \!\{\!-\infty,\!-60,-40,-30,-20,-15,-10,-8,-7,-6,-5,-4,-3,-2,-1,-1/2,\\0 ,1/4 , 1/2, 3/4, 7/8 , 1 \} $ with in each case $50$ expansion terms. For $x = 1$ and $x=-\infty$ we use a power-log ansatz, otherwise, ordinary Taylor expansions suffice. We obtain the boundary conditions in $x = 0$ again via {\tt AMFlow} with $100$ digits precision. 
\subsection{UV renormalization and IR subtraction}
\label{sec::UVIR}
The heavy-to-light form factors exhibit poles in $\epsilon$, which may be of UV- or IR-nature. Focusing on the IR subtraction,~\footnote{We refer to Ref.~\cite{Fael:2024vko} for the discussion of the UV renormalization.} we can use a universal factor $Z$~\cite{Becher:2009kw,Liu:2022elt} stemming from the SCET approach on the problem such that the finite hard matching coefficients $C$ are given by $C = Z^{-1} F$, where $F$ are the UV renormalized form factors. The matching coefficients $C$ satisfy the renormalization group (RG) equation 
\begin{equation} \label{eq:RGEC}	
	\frac{d}{d\ln(\mu)} \, C(s,\mu) = 
	\left[
	\Gamma_{\mathrm{cusp}} (\alpha_s^{(n_l)}) \, \ln\left(\frac{(1-x) m}{\mu} \right)+ \gamma^\prime(\alpha_s^{(n_l)}) + \gamma^{\mathrm{QCD}}(\alpha_s^{(n_f)}) \right] \, C(s,\mu).
\end{equation}
The cusp anomalous dimension $\Gamma_{\mathrm{cusp}}$ is known up to four loops~\cite{Henn:2019rmi,vonManteuffel:2019wbj,vonManteuffel:2020vjv,Agarwal:2021zft} and we need it to three loops. Further, we use the hard anomalous dimension $\gamma^\prime$ to three loops. It is given via $\gamma^\prime = \gamma^q+\gamma^Q$, where $\gamma^q$ $(\gamma^Q)$ is the collinear light (heavy) quark anomalous dimension. The former is calculated to four loops~\cite{vonManteuffel:2020vjv,Agarwal:2021zft}, while the latter is available up to three 
loops~\cite{Grozin:2014hna,Grozin:2015kna,Bruser:2019yjk}. $ \gamma^{\mathrm{QCD}}$ denotes the anomalous dimension of the corresponding QCD current. Additionally, $n_f$ is the number of all considered fermions ($n_f = n_l+n_h = 5$). The structure of Eq.~(\ref{eq:RGEC}) allows us to identify two scales: a scale $\mu$ governing the RG evolution in SCET and a scale $\nu$ governing the RG evolution in QCD. Therefore, we can predict the dependence of the matching coefficients $C$ on $L_\mu = \ln(\mu^2/m^2)$ and $L_\nu = \ln(\nu^2/m^2)$ from lower loop results via the running and decoupling relations for the strong coupling $\alpha_s$.
\subsection{Results}
\begin{figure}[!t]
	\centering
	\includegraphics[width=.4\textwidth]{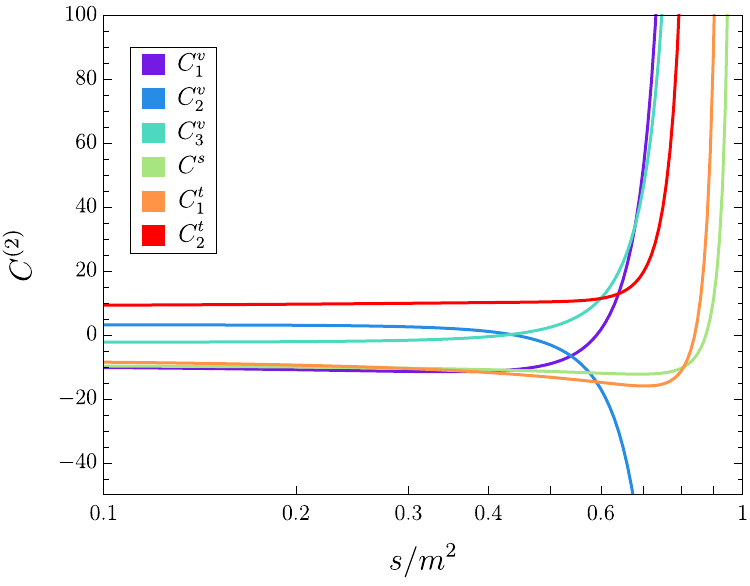}
	\hspace{0.1\textwidth}
	\includegraphics[width=.4\textwidth]{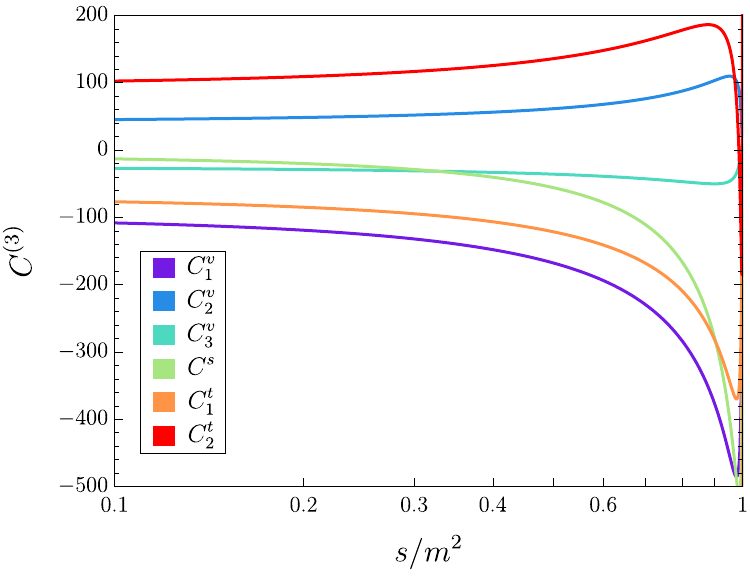}
	\caption{\label{fig::FF}Two- and three-loop hard matching coefficients $C$ as a function of $s/m^2$ for $s>0$ with colour factors adjusted to QCD ($n_l=4$, $n_h=1$). For the renormalization scale $\mu^2=m^2$ has been chosen.}
\end{figure}

All discussed results are available via the ancillary files of Ref.~\cite{Fael:2024vko} and the accompanying program {\tt FFh2l}. In Fig.~\ref{fig::FF} we show results for the two- and three-loop vector, scalar and tensor matching coefficients $C$ as a function of $s/m^2$ for $0 < s <m^2$ at $\mu^2 = m^2$. At both loop orders we observe Coulomb-like singularities close
to threshold.\\
Our results agree with Ref.~\cite{Datta:2023otd} where the colour-planar contributions (except for the tensor current) have been computed. In Ref.~\cite{Datta:2024cen} the fermionic contributions of our results for the vector, axial-vector, scalar and pseudo-scalar currents have been cross-checked and confirmed.
\section{Hard function in $\bar B \to X_s \gamma$ to three loops}\label{sec::hardbsg}
The distinction of the two scales $\mu$ and $\nu$ described in Section \ref{sec::UVIR} enables us to extract the hard function in $\bar B \to X_s \gamma$ at N$^3$LO, given via the square of a particular linear combination of the matching coefficients of the tensor current. In more general terms, the photon energy spectrum in $\bar B \to X_s \gamma$ can be written in a SCET-based approach as the product of the hard function $H$ times a convolution of a jet and a soft function~\cite{Korchemsky:1994jb,Akhoury:1995fp,Neubert:2004dd,Dehnadi:2022prz}. The hard function was up to now only known to two loops~\cite{Ali:2007sj,Ligeti:2008ac}, while the jet~\cite{Bruser:2018rad} as well as the soft function~\cite{Bruser:2019yjk} are known to three loops already. For more details on the matching procedure, we refer to Refs.~\cite{Bell:2010mg,Fael:2024vko} and references therein. For illustration, evaluating the hard function at scale $\mu = m_b$ and numerical values $C_A=3$, $C_F=4/3$, $T_F=1/2$, $n_l=4$, $n_h = 1$, the result reads
\begin{align}
	H(m_b) & = 1
	- 4.5483 \left(\!\frac{\alpha_s^{(4)}(m_b)}{\pi}\!\right) -19.2861 \left(\!\frac{\alpha_s^{(4)}(m_b)}{\pi}\!\right)^{\!\!2} -181.1617 \left(\!\frac{\alpha_s^{(4)}(m_b)}{\pi}\!\right)^{\!\!3} + \mathcal{O}(\alpha_s^4) \, .
\end{align}
Recently, we also calculated the last missing leading-colour pieces analytically, namely the (log-independent) linear-in-$n_h$ and non-fermionic terms, partially via the {\tt PSLQ} algorithm. The result reads
\begin{align}\label{eq::hardNew}	
	H_{\mathrm{LC}}(\mu) \!&=
	\Bigg(\frac{\alpha_s^{(n_l)}(\mu)}{4 \pi}\Bigg)^3
	N_c^3 \Bigg[-\Frac{277}{9} \zeta_3^2-\Frac{523391}{3265920} \pi^6+\Frac{43}{6} \zeta_5+\Frac{205}{36} \pi^2 \zeta_3 +\Frac{117311}{38880} \pi^4+\Frac{95023}{972} \zeta_3 \notag \\
	&-\Frac{24053723}{279936} \pi^2-\Frac{31613053}{52488}\Bigg] +\Bigg(\frac{\alpha_s^{(n_l)}(\mu)}{4 \pi}\Bigg)^3 N_c^2 T_F n_h
	\Bigg[6 \zeta_3^2+\Frac{103}{2835} \pi^6 + \Frac{928}{81} \pi^3 Cl_ 2 \bigg(\frac{\pi}{3}\bigg) \notag \\
	&+\Frac{2422}{9} \zeta_5 -\Frac{1054}{27} \pi^2 \zeta_3 + \Frac{632}{9} Cl_ 2^2  \bigg(\frac{\pi}{3}\bigg)-\Frac{704}{3} \mathrm{Li}_4\bigg(\frac{1}{2}\bigg)-\Frac{88}{9} \log^4 (2)+\Frac{88}{9} \pi^2 \log^2 (2) \!+\!\Frac{49}{243} \pi^4 \notag \\ 
	&+\Frac{10984}{81} \zeta_3 -\Frac{524}{27} \frac{\pi^3}{\sqrt{3}}+\Frac{15488}{27} \pi^2 \log(2)- \Frac{1394}{3} \frac{Cl_ 2 (\frac{\pi}{3})}{ \sqrt{3}}-\Frac{427787}{972} \pi^2+ \Frac{1864435}{2916}\Bigg] + \dots,
\end{align}
where $Cl_n(\theta)=\mathrm{Im} \, \mathrm{Li}_n\big(e^{i \theta}\big)$ is the Clausen function. The expression in Eq.~(\ref{eq::hardNew}) is new and reproduces the numerical results from Eq.~(50) of Ref.~\cite{Fael:2024vko}. Transcendental constants related to $\pi^3 \sqrt{3}$ and $\pi^3 Cl_2(\pi/3)$ were also observed in Ref.~\cite{schneiderLL26}. 
\section{Semileptonic $\bar B \to X_u l \bar \nu_l$ decays to three loops}
\label{sec::inclvub}
Another phenomenological application of our results is the calculation of partial decay rates in $\bar B \to X_u l \bar \nu_l$ at N$^3$LO and the resulting inclusive determination of $|V_{ub}|$. In analogy to the factorization theorem for $\bar B \to X_s \gamma$ (see Section~\ref{sec::hardbsg}), also inclusive $\bar B \to X_u l \bar \nu_l$ decays can be treated via a factorization ansatz. Again, our hard matching coefficients serve as input for the up to now unknown hard functions. In contrast to Section \ref{sec::hardbsg} we do not only determine the hard functions, but we compute the partial decay rates in $\bar B \to X_u l \bar \nu_l$ at N$^3$LO and investigate their effect on $|V_{ub}|$.\\
From the experimental point of view these decays are very challenging due to a large background from $\bar B \to X_c l \bar \nu_l$ events. For the inclusive determination of $|V_{ub}|$ cuts have to be employed to suppress the background, which have to be respected by the theoretical description of the decays as well. The genuine phase space for inclusive $\bar B \to X_u l \bar \nu_l$ decays, neglecting the mass of the lightest possible final-state pion, is given by $0 \leq P_{+} \leq P_l \leq P_{-} \leq M_B$, where $P_{\pm} = E_X \mp |\vec{P_X}|$ parametrize the hadronic system $X_u$ and $P_l = M_B - 2 E_l$, where $E_l$ is the lepton energy.  In this region of phase space, consisting of large hadronic energy and moderate hadronic invariant mass, the so-called shape-function region, the BLNP framework~\cite{Bosch:2004th,Lange:2005yw} allows for a prediction of the partial decay rates. At leading power in the $1/m_b$-expansion the BLNP expressions provide the above mentioned factorization of the decay rates into hard functions at a hard scale $\mu_h \sim m_b$, a jet function at intermediate scale $\mu_i \sim \sqrt{m_b \Lambda_{\mathrm{QCD}}}$, a shape function, describing the inner dynamics of the $B$ meson, and the possibility to consistently implement resummed higher order QCD corrections. In Ref.~\cite{Greub:2009sv} partial decay rates of  $\bar B \to X_u l \bar \nu_l$ are studied at NNLO QCD within a modified BLNP framework. As mass scheme for the $b$-quark mass $m_b$, they choose the so-called shape-function scheme, which was introduced in Ref.~\cite{Bosch:2004th}. The main result of their analysis is that, for conventional scale choices, the NNLO corrections lead to a downward shift of the partial rate, and thus (with considering also known power corrections up to $1/m_b^2$) raise the value of $|V_{ub}|$ by around $10\%$ compared to the NLO prediction~\cite{Greub:2009sv}. In this contribution we discuss the extensions of the analysis of Ref.~\cite{Greub:2009sv} to N$^3$LO at leading power.
\subsection{Factorization theorem}
\label{sec::factthm}
The central objects of our analysis are the scalar functions $f_i^{(0)}$ $(i = 1,2,3)$~\cite{Bosch:2004th,Greub:2009sv},
\begin{eqnarray}\label{eq:Resummedf}
	&& f_i^{(0)}(P_+,y)= {\rm exp}\bigg[2 S(\mu_h,\mu_i)-2S(\mu_i,\mu_0)
	- a_{\gamma^J}(\mu_i,\mu_0)-2a_{\gamma'}(\mu_h,\mu_0) 
	\nonumber \\
	&& 
	\hspace{3cm}- 2a_{\Gamma}(\mu_h,\mu_i)\ln\frac{m_b}{\mu_h}\bigg] 
	\times H_{ui}(y,m_b,\mu_h)\,y^{-2a_\Gamma(\mu_h,\mu_i)}  \\
	&&\hspace{3cm}\times
	\widetilde j\left(\ln\frac{m_by}{\mu_i}+\partial_\eta,\mu_i\right)\frac{e^{-\gamma_E\eta}}{\Gamma(\eta)}
	\int_0^{P_+}d\hat\omega\left[\frac{1}{P_+-\hat\omega}
	\left(\frac{P_+-\hat\omega}{\mu_i}\right)^\eta\right]_* \hat S(\hat
	\omega,\mu_0)\, \nonumber ,
\end{eqnarray}
where the label $(0)$ stands for the leading term in the $1/m_b$-expansion and $y = (P_{-}-P_{+})/(M_B-P_+)$. Equation~(\ref{eq:Resummedf}) shows the RG-improved factorization in terms of the hard functions $H_{ui}$, the jet function $\tilde j$ and the shape function $\hat S$, described at scale $\mu_0$. The resummation manifests itself in the RG exponents $S, a$ and $\eta$. We consistently implement into Eq.~(\ref{eq:Resummedf}) all higher-order QCD corrections up to three loops and thus can determine the  partial decay rates at the corresponding order. This means that we evaluate each of the matching functions and the RG exponents at N$^3$LO. The hard functions $H_{ui}$ are obtained from the hard matching coefficients of the vector current to three loops. Further, the jet function is calculated in Ref.~\cite{Bruser:2018rad} at this order. On the contrary, the non-perturbative shape-function $\hat S$ encodes information about the physics inside the $B$ meson and can currently not be calculated from first principles. In Section~\ref{sec::shapefunction}, we provide more details on the shape function. Phase-space integration over particular linear combinations of the functions $f_i$ respecting the employed cuts yields the corresponding value for the partial decay rate $\Gamma_u$. In the following we describe the procedure in more detail.
\subsection{Resummation ingredients}
The RG exponents are defined as~\cite{Bosch:2004th}
\begin{equation}
	\label{eq:RGexps}
	S(\nu,\mu)
	= -\int\limits_{\alpha_s(\nu)}^{\alpha_s(\mu)}\!\!d\alpha\,
	\frac{\Gamma_{\rm cusp}(\alpha)}{\beta(\alpha)} 
	\int\limits_{\alpha_s(\nu)}^{\alpha}\frac{d\alpha'}{\beta(\alpha')}
	\,, \qquad
	a_\Gamma(\nu,\mu)
	= -\int\limits_{\alpha_s(\nu)}^{\alpha_s(\mu)}\!\!d\alpha\,
	\frac{\Gamma_{\rm cusp}(\alpha)}{\beta(\alpha)} \,,
\end{equation}
and similarly for $a_{\gamma'}$ ($a_{\gamma^J}$), but with $\Gamma_{\rm cusp}$ replaced by the anomalous dimensions of the hard (jet) function. We define $\eta = 2 a_\Gamma(\mu_i,\mu_0)$. The QCD beta function has to be known through five loops~\cite{Baikov:2016tgj,Herzog:2017ohr,Luthe:2017ttg}. Similarly, we need the cusp anomalous dimension to five loops. Since it is only known to four loops we approximate the five-loop term using $\Gamma_4 = \Gamma_3^2/\Gamma_2$. The hard anomalous dimension $\gamma^\prime$ has to be provided to four loops, which is also unknown. Here, the lower-loop results suggest a poor convergence of the Pad\'e approximation. Therefore, we relate it again to the collinear light (heavy) quark anomalous dimension  $\gamma^q$ $(\gamma^Q)$ and use a Pad\'e approximation for $\gamma^Q$. Finally, the last missing piece is the four-loop jet anomalous dimension $\gamma^J$. In this case, we make use of the relation $\gamma^J = -4 \gamma^q -2 \gamma^{\mathrm{virt}}$, as recently the four-loop calculation of the virtual anomalous dimension $\gamma^{\mathrm{virt}}$ has been finalised~\cite{Kniehl:2025ttz,Gehrmann:2026qbl}. The anomalous dimensions are normalised as in Ref.~\cite{Bruser:2019yjk}.
\subsection{Shape-function scheme}
We choose the shape-function scheme for the mass of the heavy $b$-quark. The reason for using this scheme is that the heavy-quark parameters entering the partial decay rates should not only be expressed in a renormalon-free scheme, but should further respect the (bulk) properties of the shape function. In Ref.~\cite{Neubert:2004sp} the two-loop relations between heavy-quark parameters in the pole scheme and the shape-function scheme have been derived. We have to calculate these relations to one order higher in perturbation theory. In fact, it turns out that it is possible to calculate them up to four loops following the strategy of Ref.~\cite{Neubert:2004sp}. Our results will be published in Ref.~\cite{wip1:2026}. \\
In more detail, we introduce the $N$-th shape-function moment $M_N$ with a cutoff $\mu_f \gg \LambQCD$ 
\begin{align}
	M_N(\mu_f,\mu) = \int_{-\mu_f}^{\infty}\!d\omega\,\omega^N\,S(\omega,\mu),
\end{align}
where $S(\omega,\mu) = \hat S(\bar \Lambda - \hat \omega, \mu)$ and $ \bar \Lambda = m_B-m_b$. At tree-level, $M_0 = 1$ yields the normalization of the shape function, $M_1 = 0$ in the pole scheme via the equations of motion and higher moments can be expressed via heavy-quark operator matrix elements. Including higher-order corrections into $M_N(\mu_f,\mu)$ leads to a dependence on the cutoff $\mu_f$ and the renormalization scale $\mu$. Therefore, we define the running heavy-quark parameters order-by-order using the moments of the shape-function. More concretely, in the shape-function scheme we allow for a residual mass term $\delta m$ such that the first moment vanishes, i.e.\
\begin{align}
	m_b^{\mathrm{pole}} = m_b^{\mathrm{SF}}(\mu_f, \mu) + \delta m \quad \mathrm{s.t.} \quad M_1(\mu_f,\mu) = 0.
\end{align}
Additionally, we introduce the new parameter $\mu_\pi^2(\mu_f, \mu)$ in favor of the pole scheme parameter $-\lambda_1$. Since $\mu_f \gg \LambQCD$, the moments can be calculated in an operator product expansion, expressed via forward $B$-meson matrix elements of local operators. The preliminary result of the matching calculation is that 
\begin{equation}\label{MNrela}
	M_N(\mu_f,\mu) = (-1)^N \left[ \mu_f^N\,s(L,\mu)
	- N \int_{-n\cdot k}^{\mu_f}\!d\sigma\,\sigma^{N-1}\,
	s\bigg( \ln\frac{\sigma+n\cdot k}{\mu},\mu\bigg) \right] 
\end{equation}
and subsequent expansion in $n \cdot k$ with  $n\cdot k\to 0$ and $(n\cdot k)^2\to-\lambda_1/3$.\footnote{We are limiting ourselves to this order in the $1/m_b$-expansion. Further, $n$ is a light-like vector and $k$ the residual momentum of the $b$-quark inside the $B$ meson.} To obtain the scheme conversion to four loops, we need the heavy-to-light soft function $s(L,\mu)$ to four loops. From the known three-loop result~\cite{Bruser:2019yjk}, we can compute all log-dependent terms at four-loop order via its RG equation. In fact, it turns out that this is sufficient for our purposes. The only caveat is that we need the soft anomalous dimension $\gamma^{\mathrm{soft}}$ to four loops, which is unknown. For a numerical evaluation we could again resort to a Pad\'e approximation. As an important check of the scheme conversion we verify that the pole scheme parameters, expressed in the shape-function scheme, are scale-independent.
\subsection{Shape-function model}
\label{sec::shapefunction}
\begin{figure}[t]
	\centering
	
	\begin{subfigure}{0.45\textwidth}
		\centering
		\includegraphics[width=1.04\linewidth]{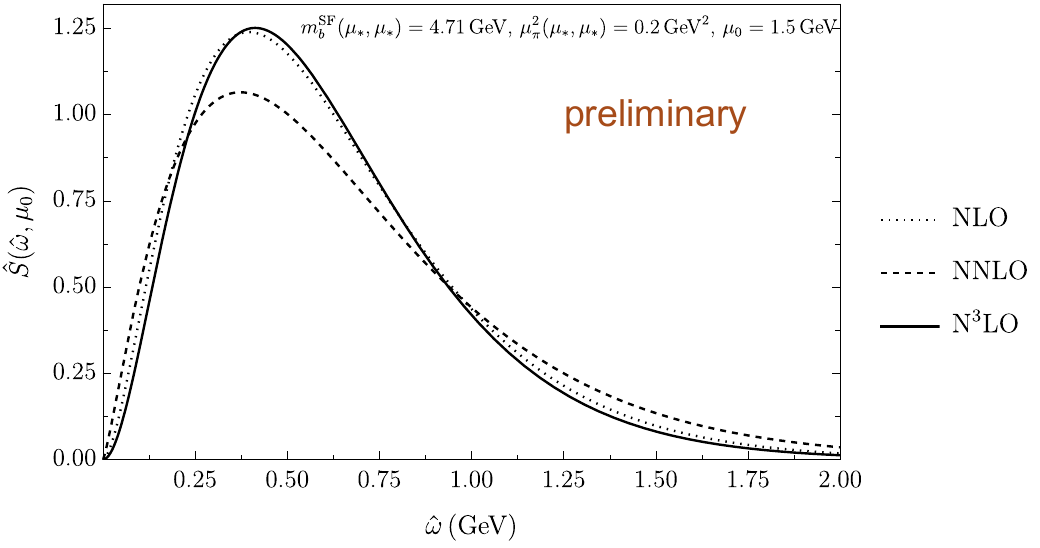}
	\end{subfigure}
	\hfill
	\begin{subfigure}{0.45\textwidth}
		\centering
		 \raisebox{.3mm}{%
		 	\includegraphics[width=\linewidth]{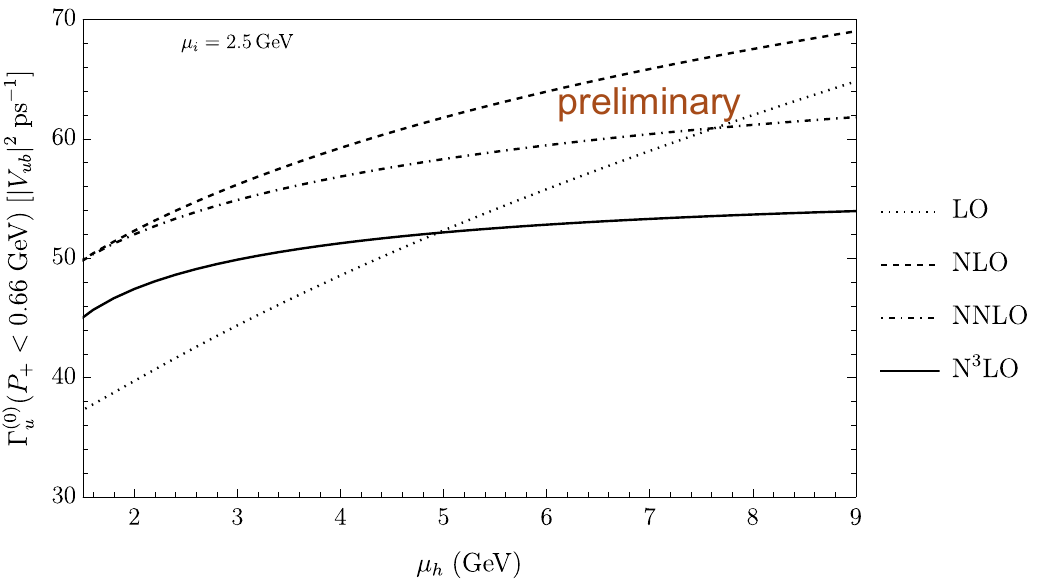}%
		 	}
	\end{subfigure}
	
	\caption{Left: Exponential model for the shape function with moment constraints evaluated at various loop orders for $m_b^{\rm SF}(\mu_\ast,\mu_\ast) = 4.71 \, {\rm GeV}$, $\mu_\pi^2(\mu_\ast,\mu_\ast) = 0.2 \, {\rm GeV}^2$. Right: Partial rates $\Gamma_u^{(0)}$ with a cut on $P_+ = 0.66 \, {\rm GeV}$ in dependence on the hard scale $\mu_h$ at intermediate scale $\mu_i = 2.5 \, {\rm GeV}$ using the exponential model for the shape-function with three-loop moment constraints at $m_b^{\rm SF}(\mu_\ast,\mu_\ast) = 4.71 \, {\rm GeV}$, $\mu_\pi^2(\mu_\ast,\mu_\ast) = 0.2 \, {\rm GeV}^2$.}
	\label{fig:sfres}
\end{figure}

Following the arguments of Section~\ref{sec::factthm} there are several ideas on how to access/model the shape function (e.g.\ Refs.~\cite{Bosch:2004th,Lange:2005yw,Ligeti:2008ac}). As a first step, we use the same exponential model as in Ref.~\cite{Greub:2009sv}, but implement the constraints on the shape-function parameters consistently with the shape-function scheme at three-loop level. The exponential model (at low scale $\mu_0$) is given via
\begin{equation}
	\label{eq:Smod}
	\hat S(\hat \omega,\mu_0)={\cal N}(b,\Lambda)\,\hat \omega^{b-1} 
	\, {\rm exp}\left(-\frac{b\hat \omega}{\Lambda}\right),  
\end{equation}
with parameters $b, \Lambda$ and normalization ${\cal N}(b,\Lambda)$. Using the running of the heavy-quark mass $m_b^{\rm SF}(\mu_f,\mu)$ and $\mu_\pi^2(\mu_f, \mu)$, together with the fact that we can obtain the moments of the shape-function directly from the model in Eq.~(\ref{eq:Smod}) by integrating up to a cutoff $\hat \omega_0$, we can construct a system of equations for $b$ and $\Lambda$ order-by-order in perturbation theory. To this end, we provide initial values for $m_b^{\rm SF}(\mu_f,\mu)$ and $\mu_\pi^2(\mu_f, \mu)$ at scale $\mu_\ast = \mu_f = \mu = 1.5\, {\rm GeV}$~\cite{Greub:2009sv,HeavyFlavorAveragingGroupHFLAV:2024ctg}. We choose the same cutoff~$\omega_0 = M_B-2E_0$ with $E_0=1.8\, {\rm GeV}$ as in Ref.~\cite{Greub:2009sv} and show the NLO, NNLO and N$^3$LO results for the shape functions in the left-hand plot of Fig.~\ref{fig:sfres}. Currently, we are implementing an improved model which contains the radiative tail of the shape function matched onto its bulk. 
\subsection{Results}
After combining all discussed pieces we can predict the decay rate $\Gamma_u$ for different cuts and scale choices. In the right-hand plot of Fig.~\ref{fig:sfres} we show results for $\Gamma_u$ with a cut on $P_+ < 0.66 \,{\rm GeV}$ while varying the hard scale $\mu_h$ and fixing the intermediate scale $\mu_i$ to a standard BLNP choice of $\mu_i = 2.5 \,{\rm GeV}$. 
We observe that the NNLO results tend to shift the value of $\Gamma_u$ downwards, reproducing the results of Ref.~\cite{Greub:2009sv}. The preliminary results of the N$^3$LO corrections to $\Gamma_u$ show a similar behaviour, i.e.\ $\Gamma_u$ is decreased further. Also the investigation of other cuts shows a similar pattern. Therefore, the N$^3$LO corrections seem to increase the value of $|V_{ub}|$, i.e.\ they also increase the gap to the exclusive determination of $|V_{ub}|$ (c.f.\ Eq.~(\ref{eq::vubExclIncl})). However, as mentioned above, these results are preliminary: We are implementing the aforementioned shape-function model incorporating a proper treatment of the tail. Furthermore, it is necessary to perform a detailed uncertainty analysis.

\section{Conclusion}\label{sec::conclusion}
In these proceedings we reviewed the calculation of the three-loop corrections of $\mathcal{O}(\alpha_s^3)$ to the heavy-to-light form factors and hard matching coefficients in SCET for generic external currents. In particular, we discussed the calculation of the master integrals and the IR subtraction. We used these results to determine the hard function in $\bar B \to X_s \gamma$ at three loops, which is now available completely analytically in the leading-colour approximation. Further, we showed first results for the N$^3$LO corrections to semileptonic $\bar B \to X_u l \bar \nu_l$ decay rates.
Preliminary studies indicate that the inclusion of the N$^3$LO corrections decreases the decay rate, and thus increases the inclusive value of $|V_{ub}|$. We stress that these investigations are ongoing and may change once the perturbative tail of the shape function and the full uncertainty analysis are included.

\section*{Acknowledgements}
We would like to thank the organisers of LL2026 for creating a very pleasant and inspiring atmosphere. Further, we are grateful to Bj\"orn O. Lange, Thomas Mannel and Benjamin~D.~Pecjak for useful discussions. The research of T.H., J.M., and M.S. was supported by Deutsche Forschungsgemeinschaft (DFG, German Research Foundation) under grant 396021762 --- TRR 257 ``Particle Physics Phenomenology after the Higgs Discovery''. T.H. and J.M. also acknowledge support from Deutsche Forschungsgemeinschaft (DFG, German Research Foundation) under Germany's Excellence Strategy~--~Cluster of Excellence ``Color meets Flavor'', EXC 3107~--~Project-ID 533766364. F.L. was supported by the Swiss National Science Foundation (SNSF) under contract \href{https://data.snf.ch/grants/grant/211209}{TMSGI2\_211209}. K.S. is supported by the European Union under the Marie Sk{\l}odowska-Curie Actions (MSCA) Grant 101204018.

\bibliographystyle{JHEP}
\bibliography{References}

\end{document}